\documentclass[prd,preprint,eqsecnum,nofootinbib,amsmath,amssymb,preprintnumbers,tightenlines,longbibliography]{revtex4-1}

\usepackage{graphicx}
\usepackage{bm}
\usepackage[colorlinks,linkcolor=blue,citecolor=blue]{hyperref}

\def\e{{\epsilon} }

\def\c{{\mathcal C}}

\def\pr{{\mathcal P}}

\def\psie{\Phi}
\def\phis{\phi }

\def\ls2{{\ell_s^2}}
\def\x{{\bm x}}
\def\dd{{\rm d}}

\newcommand\bga{\begin{align}}
\newcommand\nda{\end{align}}

\def\x{{\bm x}}

\def\p{{\bm p}}

\def\v{{\bf v}}

\def\half{{\textstyle\frac{1}{2}}}
\def\third{{\textstyle\frac{1}{3}}}

\def\nc{N_{\rm c}}

\def\st{\begin{equation}}
\def\stp{\end{equation}}
\def\bg{\begin{eqnarray}}
\def\nd{\end{eqnarray}}
\def\Eq#1{Eq.~(\ref{#1})}
\def\App#1{Appendix~\ref{#1}}
\def\Fig#1{Fig.~\ref{#1}}
\def\Sect#1{Section~\ref{#1}}
\def\Ref#1{Ref.~\cite{#1}}

\def\llangle{\left\langle}
\def\rrangle{\right\rangle}
\def\sllangle{\llangle }
\def\srrangle{\rrangle }

\def\lsim{\mbox{~{\protect\raisebox{0.4ex}{$<$}}\hspace{-1.1em}
	{\protect\raisebox{-0.6ex}{$\sim$}}~}}

\def\nott#1{\setbox0=\hbox{$#1$}                
   \dimen0=\wd0                                 
   \setbox1=\hbox{/} \dimen1=\wd1               
   \ifdim\dimen0>\dimen1                        
      \rlap{\hbox to \dimen0{\hfil/\hfil}}      
      #1                                        
   \else                                        
      \rlap{\hbox to \dimen1{\hfil$#1$\hfil}}   
      /                                         
   \fi}                                         %

\advance\parskip 1.9pt
\advance\voffset -0.2in

\def\dft{{\delta f_{(2)}}}
\def\dfo{{\delta f_{(1)}}}
\def\st{\begin{equation}}
\def\stp{\end{equation}}
\def\bg{\begin{eqnarray}}
\def\nd{\end{eqnarray}}
\def\nc{{\, ,}}
\def\np{{\, .}}

\begin{document}

\title{Second order viscous corrections to the harmonic spectrum in heavy ion collisions}

\author{D.~Teaney}
\email{derek.teaney@stonybrook.edu}
\author{L.~Yan}
\email{li.yan@stonybrook.edu}
\affiliation
    {%
    Department of Physics \& Astronomy,
    Stony Brook University,
    Stony Brook, NY 11794, USA
    }%

\begin{abstract}
   We calculate the second order viscous correction to the kinetic distribution, $\dft$, and use this result in a hydrodynamic simulation of heavy ion collisions to determine the complete second order correction to
 the harmonic spectrum, $v_n$.   At leading  order in a conformal fluid, the
 first viscous correction is determined by one scalar function, $\chi_{0p}$.
 One moment of this scalar function is constrained by the  shear viscosity.
 At second order in a conformal fluid, we find that $\delta f(\p)$ can be
 characterized two scalar functions of momentum, $\chi_{1p}$ and $\chi_{2p}$. The
 momentum dependence of these functions is largely determined by the kinematics
 of the streaming operator. Again, one moment of these functions is constrained by the
 parameters of second order hydrodynamics, $\tau_\pi$ and  $\lambda_1$. The effect
 of $\dft$ on the integrated flow is small (up to $v_4$), but is quite important for the higher harmonics at modestly-large $p_T$.  Generally, $\dft$ increases the value of $v_n$ at a given $p_T$, and is most important in small systems.
\end{abstract}

\maketitle

\section{Introduction}

Motivated by the wealth of data on collective flow in heavy ion collisions, hydrodynamic simulations of these ultra-relativistic events have steadily improved.
Today, event by event viscous hydrodynamic simulations reproduce the elliptic
fow, the triangular flow, the higher harmonic flows, and the correlations
between the flow harmonics and event plane angles \cite{[{For an up to date review, see: }] Heinz:2013th}.  
The overall agreement with the data on this
rich variety of observables 
constrains the shear viscosity of
the QGP close to transition region.
Because this success, 
there is a current need to precisely quantify the systematic uncertainties in
these simulations and the corresponding uncertainty in the extracted shear
viscosity of the QGP \cite{Luzum:2012wu}.  This paper will address this  need by
computing
the second order corrections to the thermal distribution function 
\cite{BRSSS,York:2008rr}  and by using
these results to simulate the observed harmonic spectrum.

Since the nucleus is not vastly larger than the mean free path, an important
advance in hydrodynamic simulations was the inclusion of viscous corrections to
the hydrodynamic equations of motion at first and second order in the gradient
expansion. 
In particular Baier, Romatschke, Son,  Starinets, Stephanov  (BRSSS) determined
the possible tensor forms that arise in the constituent relation of a conformal
fluid at second order \cite{BRSSS} .   These equations 
are currently used in practically all viscous simulations of heavy ion
collisions. It is satisfying that  
the gradient expansion, which underlies the hydrodynamic approach,
seems to converge \cite{Luzum:2008cw}, and seems to converge to the measured flow \cite{Heinz:2013th}.
However, 
all simulations of heavy ion collisions make additional 
kinetic assumptions about the fluid at freezeout when computing the particle spectra \cite{[{See for example: }][] Teaney:2009qa}. 
Indeed, the phase space distribution  of a viscous fluid 
$f_\p(X)$  is modified from its equilibrium form, $n_\p(X)$, by 
corrections at first and second order, $\delta f_\p = \delta f_{(1) }+ \dft$. Currently, all simulations
of heavy ion collisions compute the viscous corrections to the 
distribution function at first order, while  
using  second order corrections to the hydrodynamic equations of motion. 
The goal of this work is to remedy this inconsistency by computing
viscous corrections to the distribution function to second order.
Then, this second order correction is used to compute the harmonic spectrum,
$v_n(p_T)$. As expected \cite{Dusling:2009df},  these corrections are modest for
small harmonic numbers, $n$, and small $p_T$, but grow both with $n$ and $p_T$.

To compute the viscous corrections to the phase space distribution we will analyze kinetic
theory of a conformal gas close to equilibrium in a relaxation time approximation. This extreme idealization is still useful for several reasons. First,  a
similar equilibrium analysis of QCD kinetic theory was used to determine the second order
transport coefficients to leading order in $\alpha_s$ \cite{York:2008rr}. This analysis (which will be discussed further below) makes clear that the details of the collision
integral hardly matter in determining the second order transport coefficients.
Indeed, we will see that the structure of the second order viscous correction
$\dft$ is largely determined by the kinematics of free streaming,
rather than the details  of the collision integral.  Thus,  an analysis based
on a relaxation time approximation is an easy way to reliably estimate the size
of such second order corrections in  heavy ion collisions.  Second, the
overall normalization of $\dft$ is constrained by the second order
transport coefficients $\tau_\pi$ and $\lambda_1$, 
in much the same way that the shear viscosity constrains the normalization of $\dfo$. These  constraints
allow us to make a good estimate of the form of viscous corrections at
second order for a realistic non-conformal fluid.

With this functional form we study the effect of  $\dft$ on the harmonic spectrum in heavy ion collisions. 
In \Sect{2nddf} we  outline how $\dft$ is computed in kinetic theory. 
Then, in \Sect{implementation} we discuss the practical implementation of this formula in 
a hydrodynamic code used to simulate heavy ion collisions. These results
are used to simulate the linear response to a given deformation, $\epsilon_n$.
The linear response is a largely responsible for determining the $v_n$ in central collisions.
In non-central collisions the linear response {\it and} the quadratic response
determine the harmonic flow and its correlations \cite{Qiu:2011iv,Gardim:2011xv,Teaney:2012ke,Teaney:2012gu,Heinz:2013th}, but the quadratic
response will not be discussed in this initial study. Finally, we will summarize the effects of $\dft$ in \Sect{discussion}.

\section{2nd order corrections to the phase space distribution }
\label{2nddf}

\subsection{Notation}

Throughout the paper we will work with the metric $\eta^{\mu\nu} = (-,+,+,+)$, 
and $d=4$ notates the number of space-time dimensions.
Capital letters $P$,$X$ denote four vectors, while lower case letters 
$\p, \x$ denote three vectors in the rest frame. 

We are expanding
around a fluid in equilibrium with energy density, pressure, and flow velocity 
equal to $e(X)$, $\pr(X)$,  and  $U^{\mu}(X)$ respectively where
$U^{\mu} U_{\mu}=-1$. Thus, the rest frame projector is $\Delta^{\mu\nu} =
g^{\mu\nu} + U^{\mu} U^{\nu}$, and the spatial derivatives and temporal
derivatives are $\nabla^{\mu} = \Delta^{\mu\nu} \partial_\nu$ and $D = U^{\mu}
\partial_\mu$, respectively.  Finally bracketed tensors are rendered
symmetric-traceless and orthogonal to $U^\mu$,  
\st
\llangle A^{\mu\nu} \rrangle = \frac{1}{2} \Delta^{\mu}_{\;\rho}
\Delta^{\nu}_{\;\sigma} \left( A^{\rho\sigma} + A^{\sigma\rho} \right) -
\frac{1}{(d-1)} \Delta^{\mu\nu} \Delta_{\rho\sigma} A^{\rho\sigma}  \, .
\stp
Such tensors  transform irreducibly under rotation in the local rest frame.
A more elaborate example using $\sigma_{\mu\nu} = 2\llangle \nabla_{\mu} U_\nu \rrangle$ which appears in the algebra below is 
\begin{multline}
   \label{tensor1}
 \left\{ \sigma_{\mu_1\mu_2} \sigma_{\mu_3\mu_4} \right\}_{\rm sym}   =
\llangle \sigma_{\mu_1\mu_2}  \sigma_{\mu_3\mu_4} \rrangle 
+ \frac{4}{d + 3}\left\{ \Delta_{\mu_1\mu_2} 
\llangle \sigma^{\lambda}_{\mu_3} \sigma_{\mu_4 \lambda } \rrangle  \right\}_{\rm sym} \\
+ \frac{2}{(d-1)(d+1)} \left\{\Delta_{\mu_1\mu_2}\Delta_{\mu_3\mu_4} \right\}_{\rm sym} \sigma^2 \, ,   
\end{multline}
where the symmetrized spatial tensor is denoted with curly brackets:
\st
 \left\{ \sigma_{\mu_1\mu_2} \sigma_{\mu_3\mu_4} \right\}_{\rm sym}
 =  \frac{1}{3}\left [ \sigma_{\mu_1\mu_2} \sigma_{\mu_3\mu_4}  + \sigma_{\mu_1\mu_3}{\sigma_{\mu_3\mu_4}} + \sigma_{\mu_1\mu_4} \sigma_{\mu_2\mu_3} \right]   \, .
\stp

The equilibrium distribution 
function  is $n_\p \equiv n(-P\cdot U(X)/T(X))$ where $n(z) =  1/(e^z \mp 1)$, and $f_\p(X)$
denotes the full non-equilibrium distribution. 
The rest frame integrals are abreviated $\int_\p \equiv \int \dd^{d-1} p/(2\pi)^{d-1}$ and primes (such as  $n_p'$) denote derivatives with respect to $-P\cdot U/T$, so that $n_\p' = -n_\p(1 \pm n_\p).$  The energy and squared three momentum in the rest frame are, $E_\p = -P\cdot U$ and $p^2 = P^{\mu} P^{\nu} \Delta_{\mu\nu}$, respectively.

\subsection{Hydrodynamics}

In evaluating $\dft$ to second order we will need the 
hydrodynamic equations of motion.
In the Landau frame these equations 
read 
\st
\label{Tideal}
  T^{\mu\nu}  = e U^{\mu} U^{\nu} + \pr \Delta^{\mu\nu}  + \pi^{\mu\nu} \nc \qquad   \partial_{\mu} T^{\mu\nu} = 0  \nc
\stp
where $\pi^{\mu\nu}$ is the viscous correction to the stress tensor.
Throughout this  analysis we are working with a conformal fluid, and
consequently the bulk viscosity  is zero $\zeta=0$.
For a conformal fluid the possible tensor forms of the gradient expansion  
for $\pi^{\mu\nu}$ through second order were established by BRSSS 
\begin{multline}
\label{constituent}
\pi^{\mu\nu} = \pi^{\mu\nu}_{(1) } + \pi^{\mu\nu}_{(2)} + \ldots  = -\eta \sigma^{\mu\nu} 
 + \eta \tau_\pi \left[ \llangle D\sigma^{\mu\nu}\rrangle + \frac{1}{d-1} \sigma^{\mu\nu} \partial \cdot U \right]  \\
   +  \lambda_1 
\llangle \sigma^{\mu}_{\phantom{\mu} \lambda} \sigma^{\nu \lambda } \rrangle +  
\lambda_2  \llangle \sigma^\mu_{\phantom{\mu} \lambda} \Omega^{\nu \lambda} \rrangle + 
\lambda_3 \llangle \Omega^\mu_{\phantom{\mu} \lambda} \Omega^{\nu \lambda} \rrangle  + \ldots \nc
\end{multline}
where $-\eta\sigma^{\mu\nu} = -2\eta \llangle \nabla^\mu U^\nu \rrangle$ is the first
order term, and  the vorticity tensor is 
\st
 \Omega^{\mu\nu} = \frac{1}{2} \Delta^{\mu\alpha} \Delta^{\nu\beta} \left( \partial_\alpha U_\beta - \partial_\beta U_\alpha \right) \, . 
\stp
The ellipses in \Eq{constituent} denote terms third order in the gradients.
Using these equations of motion, the time
derivatives of the energy density and flow velocity 
can be determined from the spatial gradients of these fields
\begin{subequations}
   \label{eomtmu}
\begin{align}
De =&  -(e + \pr) \nabla \cdot U  + \frac{\eta}{2} \sigma_{\mu\nu} \sigma^{\mu\nu} + \ldots  \, , \\
DU_{\mu} =&  - \frac{\Delta_\mu \pr }{e + \pr} -  \frac{\Delta_{\mu\lambda_2}
\,  \partial_{\lambda_{1} } \pi^{\lambda_1 \lambda_2}   }{e + \pr } \,  ,
 \label{hydrop1} \\
    =& -  \frac{\nabla_{\alpha} \pr}{e + \pr}   
   + \frac{\eta}{e + \pr} \left[ (d-2) \llangle \sigma_{\mu \lambda} \nabla^{\lambda} \ln T \rrangle  + \llangle \nabla_{\lambda} \sigma^{\lambda}_{\; \mu} \rrangle \right]  + \ldots \, . \label{hydrop2}
\end{align}
\end{subequations}
In passing from  \Eq{hydrop1} to \Eq{hydrop2} we have used the first order expression 
for $\pi^{\mu\nu} = - \eta \sigma^{\mu\nu} $,  
the conformal temperature dependence of $\eta \propto T^{d-1}$, and
the lowest order equations of motion.

In hydrodynamic simulations of  heavy ion collisions  the static form
of the constituent relation \Eq{constituent} is not used. Rather,
this equation is rewritten as a dynamical equation for $\pi_{\mu\nu}$ 
which is evolved numerically \cite{BRSSS}
\begin{multline}
\label{constituent2nd}
\pi^{\mu\nu} = -\eta\sigma^{\mu\nu} - \tau_\pi \left[ \llangle D\pi^{\mu\nu} \rrangle + \frac{d}{d-1} \pi^{\mu\nu} \nabla \cdot U \right]
+ \frac{\lambda_1}{\eta^2} \llangle \pi^{\mu}_{\phantom{\mu}\lambda} \pi^{\nu\lambda} \rrangle  
- \frac{\lambda_2}{\eta} \llangle \pi^{\mu}_{\phantom{\mu}\lambda} \Omega^{\nu \lambda}   \rrangle  
+ \lambda_3 \llangle \Omega^{\mu}_{\phantom{\mu} \lambda} \Omega^{\nu\lambda} \rrangle  \, . 
\end{multline}
Similarly, when constructing $\delta f$ at first and second order we will 
systematically replace $\sigma_{\mu\nu}$ with $-\pi_{\mu\nu}/\eta$. For 
fluids with an underlying kinetic description, the transport coefficients
are additionally constrained, with $\lambda_3 =0$ and $\lambda_2 =-2\eta\tau_\pi$ \cite{York:2008rr}, 
and these relations used in the simulation.
The appropriate values for $\lambda_1$ and $\eta\tau_\pi$ will be discussed below.

\subsection{Kinetics}

To determine the viscous corrections to the distribution function we will 
solve the kinetic equations in a relaxation time approximation  through
second order in the gradient expansion 
\st
\label{Expansion}
  f_\p \equiv n_\p + \delta f_\p = n_\p  + \delta f_{(1) }  + \delta f_{(2)} + \ldots \,  .
\stp
In a relaxation 
time approximation  the Boltzmann equation reads
\st
P^{\mu} \partial_{\mu} f_\p(X)  =     -  \frac{T^2}{\c_p} \left[ f_\p(X) - n(-P\cdot U_{*}(X)/T_{*}(X)) \right] \, ,
\stp
where the dimensionless coefficient $\c_p$ is related to the canonically defined momentum dependent relaxation time
\st
 \c_{p} = T^2\,  \frac{\tau_R(E_p) }{E_p}  \, .
\stp
Following \Ref{Dusling:2009df}, we will parametrize the momentum dependence of
the relaxation time as a simple power law
\st
\label{alphadef}
\tau_{R} \propto E_\p^{1 - \alpha}\, , 
\stp
with $\alpha$ between zero and one. As we will see below,
$\alpha=0$ gives a first order a viscous correction which grows
quadratically with momentum  (which is known as the quadratic ansatz),
while $\alpha=1$ gives a first order
viscous correction which grows linearly with momentum (and is known as the linear ansatz). 

At leading order, the  parameters $T_{*}$ and $U^{\mu}_{*}$
which appear in the kinetic equation are equal 
to the Landau matched temperature and flow velocity, $T$ and $U^\mu$. 
However, starting at second 
order $T_{*}$ and $U_*^\mu$ will differ from
  $T$ and $U^{\mu}$  by  squares of gradients:
 \begin{subequations}
\begin{align}
   T_{*}(X) \equiv& T(X)  +  \delta T_*(X) \, ,  \\
   U_{*}^\mu(X) \equiv & U^\mu(X) + \delta U_*^\mu(X) \, .
\end{align}
\end{subequations}
$\delta T_{*}$ and $\delta U^{\mu}_{*}$ will be determined at each order by the 
Landau matching condition:
\[
     T^{\mu\nu} U_{\nu} = e U^{\mu} \, .
\]
Expanding $n(-P\cdot U_*/T_*)$ we have
\st
\label{nstar}
 n_\p^* \equiv n_\p +  \delta n_{p}^*  \qquad 
\delta n_{\p}^* = n_\p'\left[ -\frac{P\cdot \delta U_{*}}{T} - E_p \frac{\delta T_{*} }{T^2} \right] + \ldots \, , 
\stp
where we have used an obvious notation, $n_p^* \equiv n(-P\cdot U_*/T_*)$
and $n_\p\equiv n(-P\cdot U/T)$. 

Then to determine $\delta f$ we substitute  the expansion (\Eq{Expansion}) 
into the relaxation time equation and equate orders. 
In doing so we use the hydrodynamic equations of motion through
second  order
to write time derivatives of $T(X)$ and $U^{\mu}(X)$ in terms of 
spatial gradients of these fields. For instance, 
using  the equations of motion \Eq{eomtmu}  and
the thermodynamic identities
\begin{align}
     c_v =& \frac{de}{dT} \nc  &
     \frac{1}{T(e +\pr)} d\pr   + d\left( \frac{1}{T} \right) =& 0  \nc 
      &  c_s^2 =&  \frac{e + \pr}{T c_v} \nc  
\end{align}
and the relations $p^2 = E_p^2$  and $c_s^2 = 1/(d-1)$ of a conformal gas,
we have
\begin{multline}
\label{lhsboltz}
 P^{\mu} \partial_\mu n_\p = 
  -  n_\p' \frac{ p^{\mu} p^{\nu}} {2 T } \sigma_{\mu\nu} 
  -   \frac{n_p'}{(e + p)T}
  \Big[ -E_p P^{\mu_1} \Delta_{\mu_1\mu_2} \partial_{\lambda}\pi^{\lambda \mu_2} 
+ \half E^2_\p c_s^2 \, \eta\sigma^2  \Big] 
+ \ldots \np
\end{multline}
The term linearly proportional to  $\sigma_{\mu\nu}$ 
 is ultimately responsible for the shear viscosity, while the nonlinear terms contribute to $\dft$. 

With this discussion, we find that the $\delta f$ is determined by the hierarchy  of equations:
\begin{align}
 \label{df1sigma}
\delta f_{(1)}^{\sigma}   =&      \c_p  n_p'  \frac{P^{\mu}P^{\nu}  \sigma_{\mu\nu}  }{2 T^3}  \, , 
\end{align}
and
\begin{multline}
   \label{f2start}
   \delta f_{(2)}^{\sigma}  = 
\delta n_{\p}^*  
  + \; 
  \frac{\c_p n_\p'} {(e + p) T^3} 
  \Big[ -E_p P^{\mu_1} \Delta_{\mu_1\mu_2} \partial_{\lambda}\pi^{\lambda \mu_2} 
+ \half E^2_\p c_s^2 \, \eta\sigma^2  \Big] 
-\frac{\c_\p}{T^2} \, P^{\mu} \partial_{\mu} \delta f_{(1)}^{\sigma}     \, . 
\end{multline}
Straightforward algebra uses the equations of motion 
to decompose $\dft$ into irreducible tensors, and determines the final
form of $\delta f_{(1)}^\sigma$ and $\delta f_{(2)}^{\sigma}$ (see \App{algebra}).

We have put a superscript $\sigma$  in $f_{(1)}^{\sigma}$ and $\delta f_{(2)}^{\sigma}$ to indicate that that we are using $\sigma_{\mu\nu}$ rather than $\pi_{\mu\nu}$ 
in these equations. In realistic hydrodynamic simulations of heavy ion
collisions $\pi_{\mu\nu}$ is treated as a dynamic variable, and $-\eta\sigma_{\mu\nu}$ is systematically replaced by $\pi_{\mu\nu}$.  This yields the following reparametrization of $\delta f$
\begin{subequations}
\label{sigtopi}
\begin{align}
   \delta f_{(1)}  =& -\half \c_p n_p' \, \frac{P^{\mu}P^{\nu}}{\eta T^3} \pi_{\mu\nu}  \, ,  \\
   \delta f_{(2)}   =& \delta f_{(2) }^{\sigma} +  \half \c_p n_p' \,  \frac{P^{\mu} P^{\nu} }{\eta T^3} \left[ \pi_{\mu\nu} + \eta\sigma_{\mu\nu} \right] \, , 
\end{align}
\end{subequations}
where we have replaced $\sigma_{\mu\nu}$ with $-\pi_{\mu \nu}/\eta$ in the
first order result, and
appended the difference between these two tensors to the second order result
so that, $\delta f_{(1)} + \delta f_{(2)} =\delta f_{(1)}^{\sigma} + \delta f_{(2)}^{\sigma}$ up to third order terms.  

To record the result for $\delta f_{(2)}$, 
we first review the familiar first order case. At 
first order,  $\delta f_{(1)}$ is described by a dimensionless 
scalar function  $\chi_{0p}(E_p/T)$ 
\st 
\label{f1}
\delta f_1  =  \chi_{0p} \frac{P^{\mu_1} P^{\mu_2}} { \eta T^3} \pi_{\mu_1\mu_2} \, , 
\stp
which has been extensively studied in the litterature, and  
determines the shear viscosity \cite{Dusling:2009df}. 
In the relaxation time approximation 
this function is related to the relaxation time
\st
\chi_{0p}  =  -\half \c_p n_p' \, .
\stp
One moment of this function is constrained by the shear viscosity.
Indeed,  from the defining relation
\st
\pi^{\mu\nu}  = \int_\p  \, \frac{P^{\mu} P^{\nu}}{P^0} \, \delta f_\p \,  ,
\stp
we determine the shear viscosity
\st
\eta = \frac{2}{(d-1)(d + 1) T^3 } \int_\p  \, \frac{p^4}{E_p}   \chi_{0p} \, ,
\label{sheareq}
\stp
and a constraint on $\dft$
\st
\label{constrain2}
0 = \int_\p  \frac{P^{\mu} P^{\nu} }{P^0}  \dft\, .
\stp
This constraint reflects the reparametrization of $\sigma^{\mu\nu}$ in the first order $\dfo$ with $\pi^{\mu\nu}$. 
For later use and comparison, we note that the enthalpy is 
\st
\label{enthalpy}
(e + \pr)  =  \frac{-1}{(d-1) T} \int_\p \,  n_p'  \, p^2 \, , 
\stp
which can be obtained by comparing the stress tensor from kinetic
theory for small fluid velocities ({\it i.e.} $U^{\mu} \simeq (1, \v)$ with $\v \ll 1$)  to ideal hydrodynamics,  $T^{0i} \simeq (e + \pr) v^{i}$ 
\cite{Teaney:2006nc}. 

At second order the function  $\dft$ is described 
by two dimensionless scalar functions $\chi_{1p}$ and $\chi_{2p}$
\begin{align}
   \chi_{1p} =&  -\half \c_p  \, \chi_{0p}' \, ,  \\
   \chi_{2p} =&  \c_p \,  \chi_{0p} \, .
\end{align}
Two moments of these scalar functions are constrained by the second order transport coefficients
$\eta \tau_\pi$ and $\lambda_1$
\begin{align}
   \lambda_1 + \eta \tau_\pi =&  \frac{8}{(d-1)(d+1) (d+3) T^6}  
   \int_\p \,  \chi_{1p} \, \frac{p^6}{E_p}  \, , 
   \label{l1eq}
   \\
   \eta\tau_\pi =&  \frac{2}{(d - 1)(d+1) T^5} \int_\p
   \,  \chi_{2p} \, p^4  \, .
   \label{nteq}
\end{align}
Then the functional form of $\dft$ is 

\begin{align}
\label{f2}
\dft=&
   \frac{\chi_{1p} }{\eta^2}  \, \frac{P^{\mu_1}P^{\mu_2}P^{\mu_3}P^{\mu_4} }{T^6} \llangle \pi_{\mu_1\mu_2} \pi_{\mu_3 \mu_4} \rrangle 
   +  \frac{\chi_{2p}} {\eta} \,\frac{P^{\mu_1} P^{\mu_2} P^{\mu_3}}{T^5} 
   \left[   (d+2) \llangle \pi_{\mu_1\mu_2} \nabla_{\mu_3} \ln T \rrangle 
   - \llangle \nabla_{\mu_1} \pi_{\mu_2 \mu_3}  \rrangle  
\right] \nonumber \\ 
  &\quad  + 
   \frac{\xi_{1p}}{\eta^2} \, \frac{P^{\mu_2} P^{\mu_1}  }{T^4}
 \llangle \pi^\lambda_{\;\mu_2}\pi_{\mu_1\lambda} \rrangle  
  + \frac{\xi_{2p}}{\eta }   \frac{P^{\mu_2} P^{\mu_1}}{T^3} \left[ 
\pi_{\mu_2\mu_1} + \eta\sigma_{\mu_2\mu_1} \right] +   \nonumber 
 \frac{\xi_{3p}}{\eta}
 \frac{P^{\mu_2}}{T^3} \left[ 
       \Delta_{\mu_2\lambda_2} \partial_{\lambda_1} \pi^{\lambda_1 \lambda_2}   
    \right] \nonumber \\ &  
       \quad +  \frac{\xi_{4p}}{T^2 \eta^2} \pi^2 \, , 
\end{align}
where the four scalar functions $\xi_{1p},\xi_{2p},\xi_{3p},\xi_{4p}$ 
are linearly related to  $\chi_{0p},\chi_{1p}, \chi_{2p}$
\begin{subequations}
\label{xieq}
\begin{align}
   \xi_{1p} &= \chi_{1p} \, \frac{4 \bar p^2}{(d+3)} - \frac{\chi_{2p}\, \bar E_p }{\eta\tau_\pi} (\eta\tau_\pi+ \lambda_1) \, ,  \\
   \xi_{2p} &=  \frac{\chi_{2p}}{T\tau_\pi} \, \bar E_p   - \chi_{0p}   \, , \\
   \xi_{3p} &=  -\chi_{2p} \frac{2 \bar p^2}{(d+1)}  + 2 \chi_{0p} \frac{\eta}{s} \bar E_p    +  a_{P_*} n_p' \, ,  \\
   \xi_{4p} &=  \chi_{1p} \frac{2\bar p^4}{(d-1)(d+1) } - \chi_{2p} \frac{\bar E_p \bar p^2}{(d-1)} - \chi_{0p} \frac{\eta}{s} \bar E^2_p c_s^2 + a_{E_*}  n_p' \bar E_p \, ,
\end{align}
\end{subequations}
with $\bar{p} = p/T$ and $\bar{E}_\p = E_\p/T$.

The coefficients $a_{E*}$ and $a_{p*}$   come from \Eq{nstar}  and 
are adjusted so that the Landau matching conditions are satisfied. 
More specifically, we choose   $\delta U_{*}$ and $\delta T_{*}$ in \Eq{nstar}  so
that
\begin{subequations}
\begin{align}
   - \frac{P\cdot \delta U_{*}}{T} =&   a_{P_*} \, \frac{P^{\mu_1}}{\eta T^3} \left[ \Delta_{\mu_1 \lambda_2} \partial_{\lambda_1} \pi^{\lambda_1 \lambda_2}  \right] \, ,  \\
-E_p \frac{\delta T_{*}}{T^2}  =&  a_{E_*} \, \bar{E}_p \, \frac{\pi^2}{T^2\eta^2} \, .
\end{align}
\end{subequations}
Then integrating over  $f_\p(X)$  to determine the stress tensor
and demanding that \Eq{constrain2} (which is a restatement of the Landau matching condition),  we conclude that
\begin{subequations}
\begin{align}
   a_{P_*} =& \frac{T \eta\tau_\pi}{s}  - (1+d) \left(\frac{\eta}{s}\right)^2   \, , \\
   a_{E_*} =& \frac{T \eta\tau_\pi}{4 s}  
   - \frac{d+3}{d-1}\,  \frac{T\lambda_1}{4s} +\frac{d+1}{2(d-1)} \, \left(\frac{\eta}{s}\right)^2 \, .
\end{align}
\end{subequations}


Despite being somewhat complicated, the functional
form of $\dft$
is  severely constrained, and is bounded by the transport coefficients $\eta$, $\lambda_1$
and $\eta \tau_\pi$  through Eqs.~(\ref{sheareq}), (\ref{l1eq}), and (\ref{nteq}). 
For a single component classical gas with the quadratic ansatz $\alpha=0$, \Eq{sheareq} 
shows that $\c_p=\frac{\eta}{s}$, and the three scalar functions which determine $\dft$ can be simplified to 
\begin{align}
   \chi_{0p} = \frac{\eta}{2s} n_p \,,  \qquad \chi_{1p} = \frac{1}{4} \left(\frac{\eta}{s} \right)^2  n_p \, , \qquad \chi_{2p} = \frac{1}{2} \left(\frac{\eta}{s} \right)^2 n_p   \,. 
\end{align}
We will discuss the implementation of $\dft$ in the next section.

\section{Implementation in simulations of heavy ion collisions}
\label{implementation}

In this section we will implement the $\dft$ corrections
in a 2+1 boost invariant hydrodynamic code.   A full event-by-event simulation
of heavy ion collisions with $\dft$, together with a comparison
to data, goes beyond the scope of this initial study.  Nevertheless, 
the effect of $\dft$ in larger simulations can be anticipated 
by understanding how the linear response is modified by $\dft$.
Indeed, the qualitative features 
of event-by-event hydrodynamic simulations of heavy ion collisions 
(including the correlations between the
harmonics of different order) 
are reproduced by 
linear {\it and} quadratic response \cite{Qiu:2011iv,Gardim:2011xv,Teaney:2012ke}. In central collisions
the linear response is sufficient, and was recently used to
produce one of the best estimates of the shear viscosity and its uncertainty to
date \cite{Luzum:2012wu}.  We will calculate the linear response to 
a given deformation $\epsilon_n$ in order to estimate the influence of $\dft$ on $v_n$.

In a given heavy ion event, the particle spectrum  is expanded in
harmonics
\st
\label{definition}
\frac{1}{p_T} \frac{dN}{dp_T d\phi_\p} = \frac{1}{2\pi p_T} \frac{dN}{dp_T} \left( 1 +  \sum_n v_n(p_T) e^{in(\phi_\p - \Psi_n(p_T)) }  +  \mbox{complex conj.} \right) \, , 
\stp 
where $\phi_\p$ is the azimuthal angle around the beam pipe, and
$v_n(p_T)$ are positive by definition.
In a linear response approximation  the $n$-th harmonic, $v_n(p_T)$,  in the event is assumed to be proportional
to the $n$-th cumulant, $\epsilon_n$,  which characterizes the  deformation
of the entropy distribution\footnote{Note that we will use cumulants rather than
moments to characterize the deformation \cite{Teaney:2010vd,Teaney:2012ke}.}.  
Specifically,  we first define the normalized entropy distribution  at an time initial time $\tau_0$  
\st
\rho(\x_\perp)   \equiv \frac{\tau_0 s(\x_\perp)}{\int \dd^2x\, \tau_0 s(\x_\perp)} \, .
\stp 
Writing the coordinates in the transverse plane $\x_\perp= (x,y)$  as 
a complex number,   $z = x + iy = r e^{i\phis}$,    we define
the first six cumulants characterizing the harmonic deformations 
of the initial distribution 
\begin{subequations}
\begin{align}
   \e_1 e^{i \psie_1}  \equiv&   -\frac{\sllangle z^{*}z^2 \srrangle }{\sllangle r^3 \srrangle } \, ,  \\
   \e_2 e^{i 2\psie_2}  \equiv&  -\frac{\sllangle z^2 \srrangle  }{\sllangle r^2 \srrangle }  \, , \\
 \e_3 e^{i3\psie_3}  \equiv& -\frac{\sllangle z^3 \srrangle }{ \sllangle r^3 \srrangle  } \, ,  \\
   \e_4 e^{i 4\psie_4 }  \equiv&  \frac{1}{\llangle r^4 \rrangle} \left[ \llangle z^4 \rrangle - 3 \llangle z^2  \rrangle^2  \right] \, , \\
   \e_5 e^{i 5\psie_5 }  \equiv&  \frac{1}{\llangle r^5 \rrangle} \left[ \llangle z^5 \rrangle - 10 \llangle z^2  \rrangle \llangle z^3 \rrangle  \right]  \, , \\
   \e_6 e^{i 6\psie_6 }  \equiv&  \frac{1}{\llangle r^6 \rrangle} \left[ \llangle z^6 \rrangle - 15 \llangle z^4  \rrangle \llangle z^2 \rrangle  - 10 \llangle z^3 \rrangle^2 + 30 \llangle z^2 \rrangle^3 \right]  \, ,
\end{align}
\end{subequations}
where $\llangle \ldots \rrangle$ denote an average over  $\rho(\x_\perp)$. 

In 
a linear response approximation the orientation angle of $n$-th harmonic  $\Psi_n(p_T)$ is aligned  or anti-aligned with the cumulant angle $\Phi_n$. 
Specifically, the spectrum in linear response is \cite{Teaney:2010vd}  
\st
\label{linresponse}
\frac{1}{p_T} \frac{dN}{dp_T d\phi_p} = \frac{1}{2\pi p_T} \frac{dN}{dp_T} \left(1 + 2\sum_n w_{n}(p_T)  \cos(n(\phi_\p - \psie_n))  + \ldots \right) \, . 
\stp
Comparison with \Eq{definition}  
shows that within this approximation scheme
\st
v_{n}(p_T) = \sqrt{w_{n}^2(p_T)} \, , 
\stp
and thus $w_n(p_T)$ differs at most by a sign from $v_n(p_T)$, {\it i.e.} $w_n(p_T) = v_n(p_T) \cos(n(\Psi_n(p_T) - \psie_n))$. We will present the linear response coefficient, $w_n(p_T)/\epsilon_n$.

The linear response coefficient $w_{n}(p_T)/\epsilon_n$  is independent of 
many of the details of the initial state \cite{Teaney:2010vd}, 
and can be reasonably computed by initializing
$2+1$ boost invariant hydrodynamics with a deformed Gaussian distribution, 
where  the rms radius  and amplitude of the Gaussian are adjusted to match
the rms radius and total entropy of the event.  For example, to simulate $w_3$
at RHIC at an impact parameter of $b=7.45\,{\rm fm}$ we initialize a Gaussian
deformed by $\epsilon_3=0.05$ with $\Phi_3 = 0$
\st
\label{triangle_space}
\tau_o s(\x,\tau_o)   = C_s \llangle N_\p \rrangle \left[1 +  \frac{\llangle r^3 \rrangle \epsilon_3}{24}  \left( \left(\frac{\partial }{\partial x} \right)^3 - 3 \left(\frac{\partial}{\partial y}  \right)^2 \frac{\partial}{\partial x} \right) \right]  \frac{e^{- \frac{r^2}{\llangle r^2 \rrangle } } }{\pi \llangle r^2 \rrangle }
 \, , 
\stp
where $C_s$ sets the total multiplicity in the event. 
Here $\llangle r^2 \rrangle$ and $\llangle N_\p \rrangle$ are 
computed  using the Phobos Glauber model \cite{Alver:2008aq}.
In order that the 
total entropy closely matches the total entropy in more complete simulations
\cite{Luzum:2008cw,Luzum:2010ag}, we take $C_s = 15.9$ and  $28.04$ at RHIC and
LHC respectively.  This procedure has been used previously by the authors to
determine the linear and non-linear response \cite{Teaney:2010vd,Teaney:2012ke, Teaney:2012gu}. 

After initializing  the  Gaussian, we evolve the system  
with second order hydrodynamics, \Eq{Tideal} and \Eq{constituent2nd}, 
using a variant of the central scheme  developed previously  \cite{Pareschi,Dusling:2007gi}.
Then for a given $n$-th order harmonic perturbation $\epsilon_n$ we compute 
$w_n/\epsilon_n$ by performing the freezeout integral at a constant
temperature. 
This evolution requires
an equation of state and specified hydrodynamic parameters 
at first and second order.  In what follows we will consider a
conformal equation of state for a single component classical 
gas, and a lattice motivated equation of state previously
used by Romatschke and Luzum \cite{Luzum:2008cw}. 

Since it is only for the conformal  equation of state $p=\third\epsilon$
that the analysis of \Sect{2nddf} is strictly valid we will discuss this case
first, and then discuss the necessary modifications for a lattice based equation of state. 
To keep the final freezeout volume of the conformal equation of state approximately  equal to the much more realistic lattice based equation of state, we choose the final freezeout temperature ($T_{\rm fo}=96\, {\rm MeV}$) so that the entropy density at freezeout $s_{\rm frz}=1.87\,{\rm fm}^{-3}$
equals the entropy density of a hadron resonance gas at a temperature of
$T=150\,{\rm MeV}$. The relation
between the temperature and energy density for the conformal equation of state is $e/T^4=12.2$, which 
is the value for a two flavor ideal quark-gluon plasma. The motivation for these choices,  the parameters  of the
conformal equation of state,  and  further details about the initial conditions
and freezeout we refer to our previous work  -- see especially Appendix B of \Ref{Teaney:2010vd}.

The second order transport coefficients $\eta\tau_\pi$ and
$\lambda_1$ are all constrained by 
the  momentum dependence of the relaxation time and the
shear viscosity through Eqs.~(\ref{sheareq}), (\ref{l1eq}), (\ref{nteq}). As discussed
in \Ref{Dusling:2009df}, there are two limits for this momentum dependence
which span the gamut of reasonable possibilities. In the first
limit the relaxation time grows  linearly with momentum, and $\alpha=0$
in \Eq{alphadef}. This is known as the {\it quadratic ansatz}, and is most
often used to simulate heavy ion collisions. In a similarly extreme limit
the relaxation time is independent of momentum, and $\alpha = 1$ in 
\Eq{alphadef}. This is known as the {\it linear ansatz}, and this limit provides
a useful foil to the more commonly adopted quadratic ansatz. Once the 
shear viscosity and the momentum dependence of the relaxation time are
given, the  collision kernel is completely specified in the relaxation time
approximation, and all transport coefficients are fixed.
For a linear and quadratic ansatz 
we record the appropriate second order transport coefficients in Table~\ref{second_order_table}.
\begin{table}
   \begin{center}
      \renewcommand\arraystretch{1.5}
      \renewcommand\tabcolsep{12pt}
      \begin{tabular}{c|c|c}
         Momentum Dependence & $\eta \tau_\pi$ &  $\lambda_1$  \\ \hline
         Linear Ansatz ($\alpha=1$) &  $(d+1)=5$   & $(d+1)^2/(d+3)=25/7$\\
         Quadratic Ansatz ($\alpha=0$) & $(d+2)=6$ &  $(d+2)=6$ \\
      \end{tabular}
      \caption{A compilation of rescaled second order transport coefficients for
         a linear and quadratic ansatz in a relaxation time approximation for
         classical statistics \cite{York:2008rr}. All numbers in this table
         should be {\it multiplied} by $\eta^2/(e + \pr)$.  In a relaxation
         time approximation $\lambda_2 = - 2\eta \tau_\pi$ and $\lambda_3 = 0$
         \cite{York:2008rr}.
      \label{second_order_table}  }
   \end{center}
\end{table}

\begin{figure}
   \includegraphics[width=0.49\textwidth]{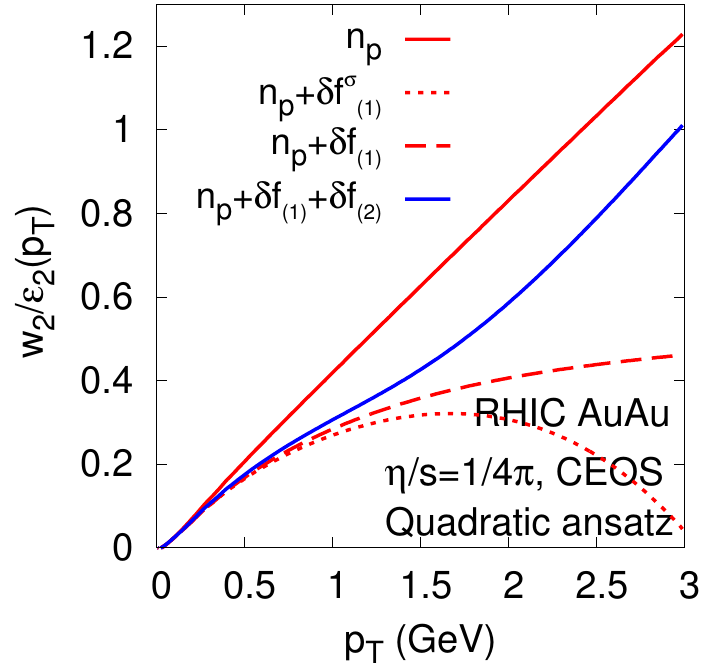} 
   \includegraphics[width=0.49\textwidth]{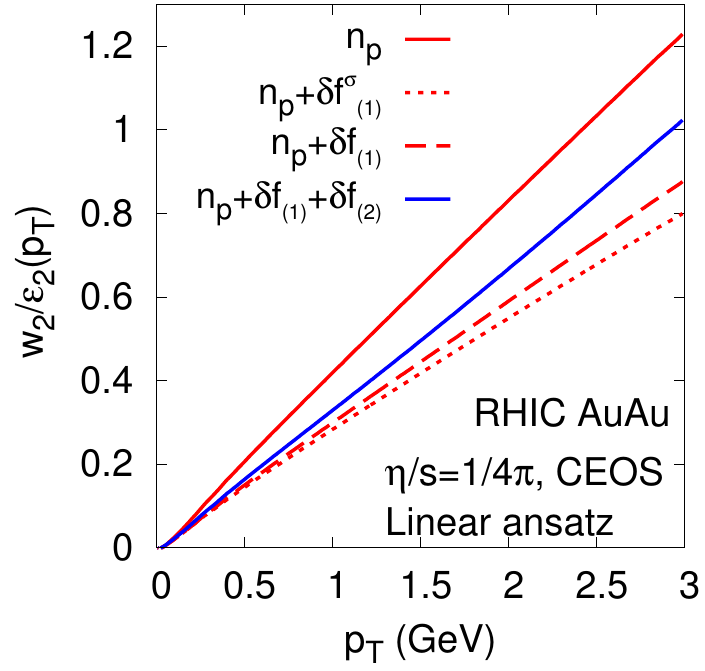} 
   \includegraphics[width=0.49\textwidth]{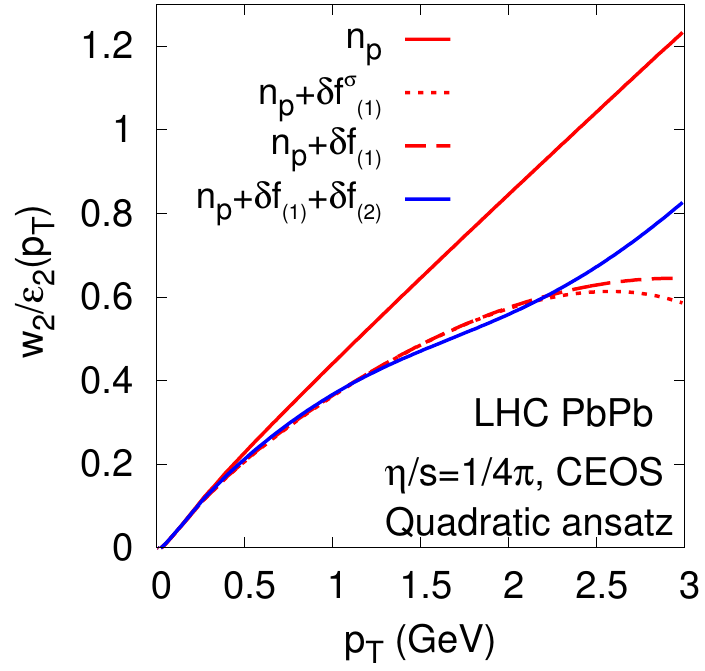} 
   \includegraphics[width=0.49\textwidth]{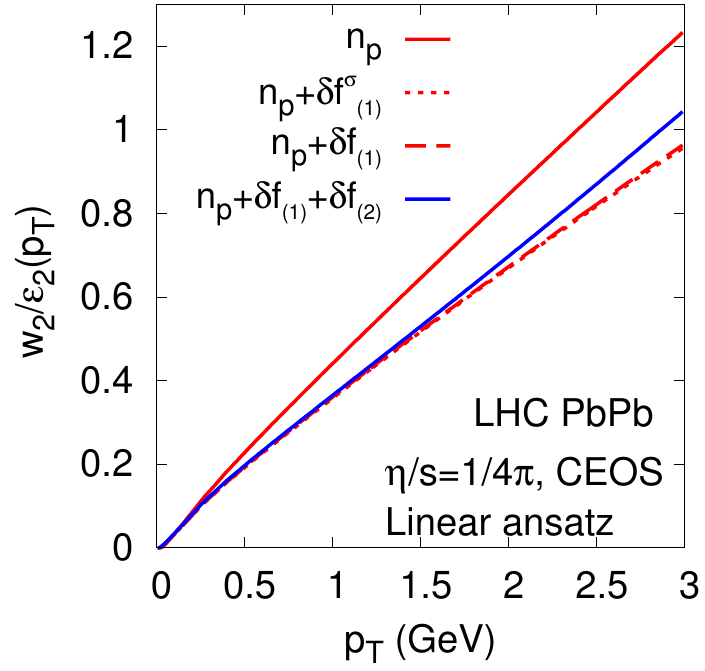} 
   \caption{
      Differential $w_2(p_T)/\epsilon_2$  for RHIC 
      and LHC initial conditions,  for a linear and quadratic ansatz, 
      and a conformal equation  equation of state (CEOS).  
      The $n_p$ curves shows the flow from second order hydrodynamics without
      the viscous correction to the distribution function;  $\delta f_{(1)}$
      and $\delta f_{(1)}+ \delta f_{(2)}$ show the flow with the viscous correction at first and second order; and finally,
      the $\delta f_{(1)}^\sigma$ result  uses $-\eta\sigma^{\mu\nu}$ instead of $\pi^{\mu\nu}$ in the first order result (see \Eq{df1sigma}).
      The freezeout
      temperature is chosen so that the freezeout entropy density  
      of the conformal gas equals that of a hadronic resonance gas 
      at a temperature of $T=150\,{\rm MeV}$. 
\label{v2ceos}
}
\end{figure}

So far this section has detailed the initial and freezeout conditions, 
as well as the second order parameters which are used in the conformal
equation of state.
\Fig{v2ceos}  shows the resulting elliptic flow  
for the quadratic and linear ans\"atze  for a conformal equation 
of state including the first and second order $\delta f$.   
A conformal equation of state
has a strong expansion, and, as a result, generally over estimates the
magnitude of the $\delta f$ correction. Thus the conformal analysis provides a
schematic upper bound on the magnitude of the $\dft$  correction. Further
discussion of these results is reserved for \Sect{discussion} 


Strictly speaking the
analysis of \Sect{2nddf} is useful only for a single component conformal
gas.
Nevertheless, we believe the usefulness of the analysis extends
beyond this limitted regime \cite{York:2008rr}.  Indeed, examining the steps in 
the derivation  one finds that only very-few non conformal terms 
appear at each order.   For instance, if non-conformal corrections 
are kept in \Eq{df1sigma} one finds
\st
\delta f_{(1)-{\rm non-conf}}^{\sigma}(\p) =  \c_p n_p' \left[ \frac{P^{\mu}P^{\nu} \sigma_{\mu\nu}}{2T^3} +  \left( -\frac{E_\p^2 - p^2}{3T^3} + \frac{(\third - c_s^2) E_\p^2 }{T^3} \right) \nabla_{\mu} U^{\mu}  \right] \, , 
\label{df1noncfrm}
\stp
which shows that non-conformal terms (the second term in \Eq{df1noncfrm}) are either
suppressed by $\third - c_s^2$, or are suppressed 
at  high momentum relative to the conformal
terms. 

To extend our analysis to a multi-component non-conformal equation of state we have followed the simplified treatment that is used in almost all 
simulations of heavy ion collisions. First,  we will treat  all species independently
\st
P^{\mu} \partial_{\mu} f_\p^a(X)  =     -  \frac{T^2}{\c_p^a} \left[ f^a_\p(X) - n^a(-P\cdot U_{*}(X)/T_{*}(X)) \right] \, ,
\stp
where $a=\pi,K,\rho, \ldots $ is a species label\footnote{There have been several efforts to go beyond this extreme species independent approximation 
\cite{Dusling:2009df,Denicol:2012yr}.
}.
We will also adopt the quadratic ansatz $\alpha=0$, so that 
$\c_p^a$ is independent of momentum.  
Then for every species we define the partial  entropy 
and shear viscosity as in Eqs.~\ref{sheareq} and \ref{enthalpy} 
\begin{subequations}
\begin{align}
   \eta_a =& \frac{-\c_p^a}{(d-1)(d + 1) }  \left[ \frac{g_a}{T^3} \int_\p  {n_p^{a}}' \, \frac{p^4}{E_p}  \right] \, , \\
   s_a  =&  \frac{-1}{(d-1) } \left[ \frac{g_a}{T^2} \int_\p \,  {n_p^{a}}'  \, p^2 \right] \, , 
\end{align}
\end{subequations}
where $g_a$ is the spin-isospin degeneracy factor. The full shear viscosity
and entropy density is a sum of the partial results,  $\eta = \sum_a \eta_a$ and $s = \sum_a s_a$.
We require that $\eta_{a}/s_{a}$ is equal to $\eta/s$ for each
species, and thus the relaxation time parameter $\c_p^a$ is, in principle, 
different for each species.
However, for a classical gas the two integrals in square 
brackets are equal upon 
integrating by parts, and thus
$\c_p^a = \eta_a/s_a= \eta/s$ is indepenent
of the mass and species label. For fermi-dirac and bose statistics
these integrals are very nearly equal (to 4\% accuracy) for all values
of the mass, and $\c_\p^a$ is approximately equal to $\eta/s$ for all
species independent of mass and statistics.

Now that the relaxation time parameter $\c_p^a$ is fixed  for 
each species, the corresponding second order $\delta f$  for each
species is found by appending a species label, $n_p \rightarrow n_p^a$ and $\c_p\rightarrow \c_p^a$, to previous results.
For a multi-component gas with a quadratic ansatz 
we find
\begin{subequations}
\begin{align}
   \eta \tau_\pi + \lambda_1  =&  \sum_a (\c_p^a)^2  \left[ \frac{ 2g_a}{(d-1)(d+1)(d+3) T^6} \int_\p  {n_p^a}'' \, \frac{p^6}{E_p} \right] \, , \label{multi1}\\
   \eta \tau_\pi  =&  \sum_a (\c_p^a)^2  \left[ \frac{-g_a}{(d-1)(d+1) T^5} \int_\p {n_p^a}' \, p^4 \right] \, .
\end{align}
\end{subequations}
For classical statistics $\c_p^a=\eta/s$, and integrating
the first integral in square brackets  by parts 
yields
a simple relation noted previously \cite{York:2008rr}   
\begin{align}
\lambda_1 =\eta\tau_\pi  \qquad \qquad\mbox{(for $\alpha=0$)}  \, .
\end{align}
The remaining thermodynamic integrals  are most easily done numerically;
summing over all hadronic species  with mass less than $1.5\,{\rm GeV}$ we 
find
\begin{align}
   \lambda_1 = \eta\tau_\pi =&  \frac{\eta^2}{(e+\pr)} 8.9 \, . 
\end{align}
Thus,  $T\tau_\pi/(\eta/s) = 8.9$ would seem to be the most consistent value for 
the 2nd order transport coefficients during the hydrodynamic evolution of the hadronic phase.  However, this value
for $T\tau_\pi$ is somewhat too large to be used comfortably in the simulation
\cite{[{See for example: }][{}]Song:2007ux}.
Further,  $T\tau_\pi$ decreases  as temperature increases,
and $T\tau_\pi/(\eta/s) \simeq 5$ is  good
approximation in the QGP phase \cite{York:2008rr}.  We have therefore taken $\lambda_1 = \eta\tau_\pi = 5 \eta^2/(e + \pr)$ throughout the evolution. This means that there is a small inconsistency between the second order $\delta f$ at
freezeout, and the second order parameters used to simulate the bulk of the
hydrodynamic evolution.  
Similar inconsistencies are found in all attempts to consistently 
couple hydronamic codes with hadronic cascades \cite{Song:2011hk}.

\begin{figure}
   \includegraphics[width=0.49\textwidth,height=7.6in]{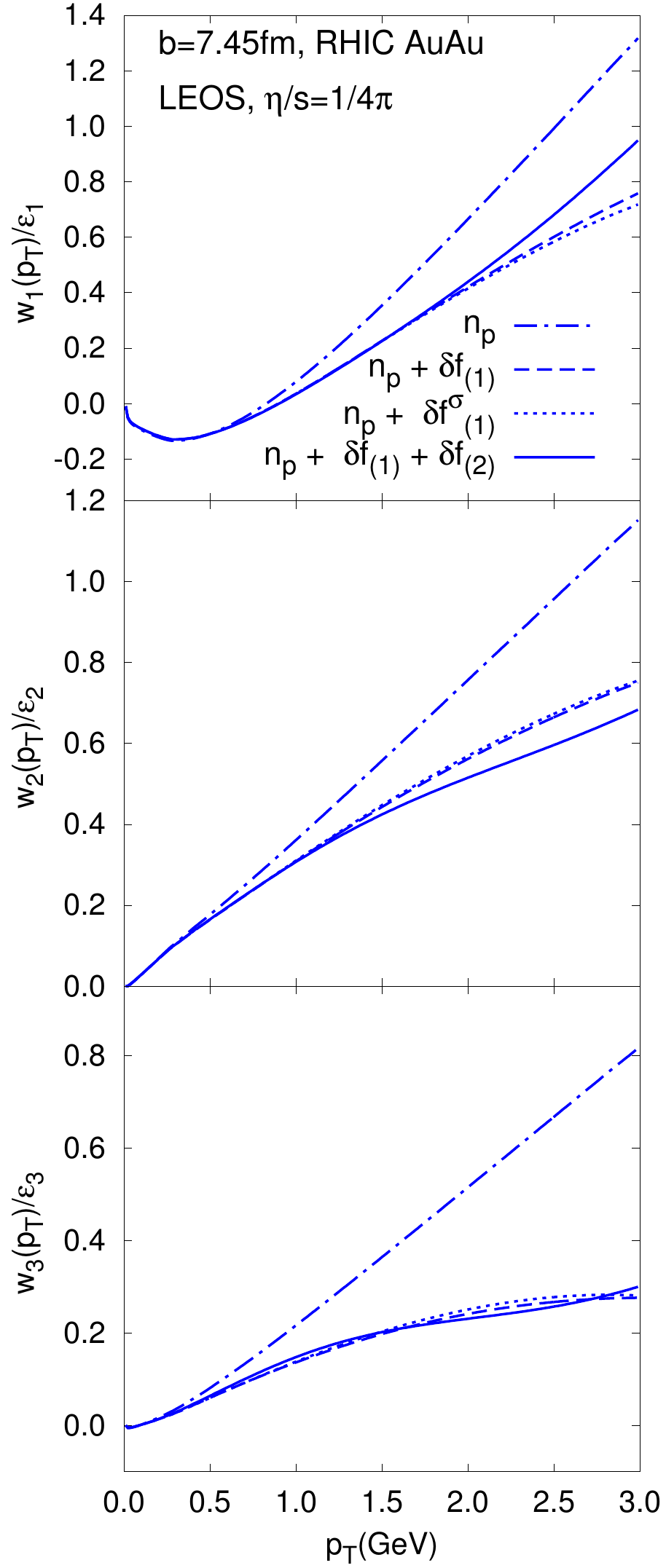} 
   \includegraphics[width=0.49\textwidth,height=7.6in]{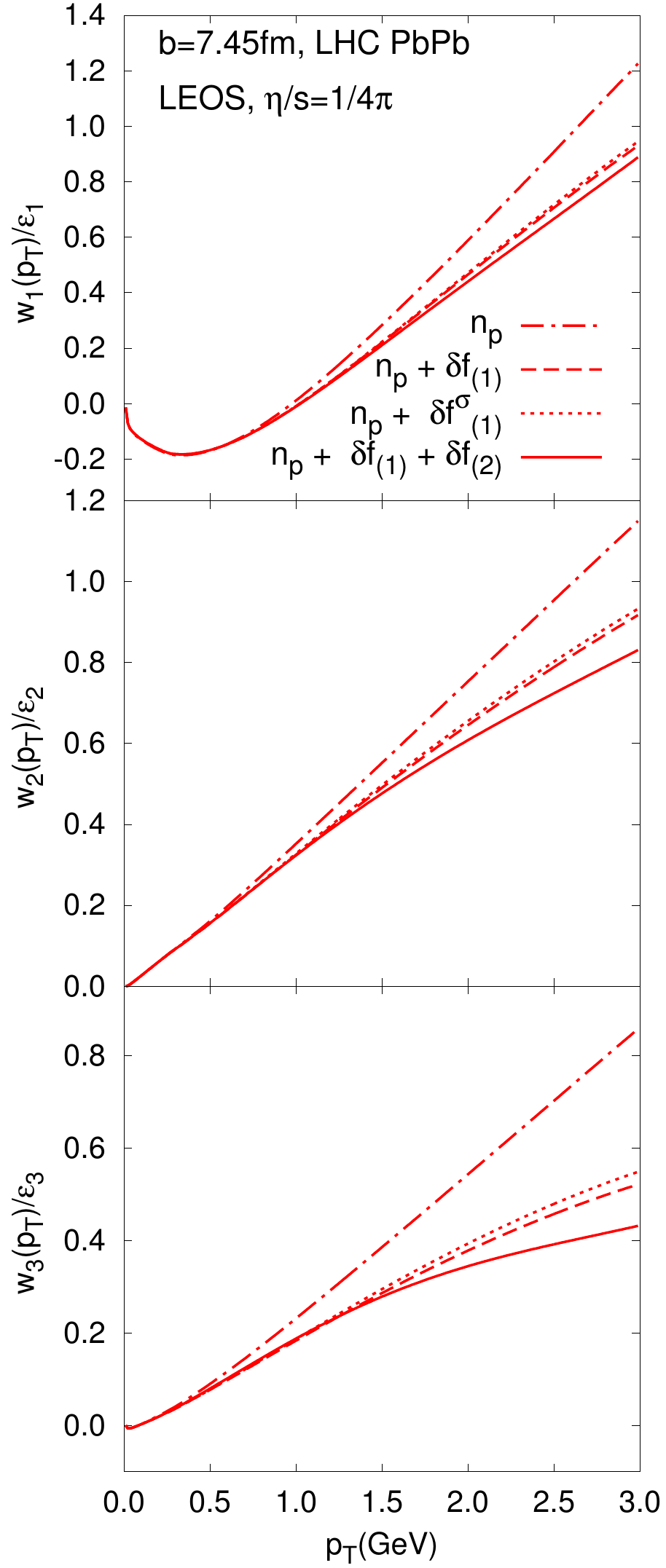} 
   \caption{Differential $w_n/\epsilon_n$   from various viscous
      hydrodynamic simulations  at $b=7.45\,{\rm fm}$ for RHIC and LHC initial conditions, and a lattice equation of state (LEOS) with $T_{\rm fo}=150\,{\rm MeV}$.
      Here 
      the $n_p$ curve shows the flow from second order hydrodynamics without
      the viscous correction to the distribution function;  $\delta f_{(1)}$
      and $\delta f_{(1)}+ \delta f_{(2)}$ show the flow with the viscous correction at first and second order respectively; and finally,
      the $\delta f_{(1)}^\sigma$ curve  uses $-\eta\sigma^{\mu\nu}$ instead of $\pi^{\mu\nu}$ in the first order viscous correction (see \Eq{df1sigma}).
\label{v1tov3} 
}
\end{figure}

\begin{figure}
   \includegraphics[width=0.49\textwidth,height=7.6in]{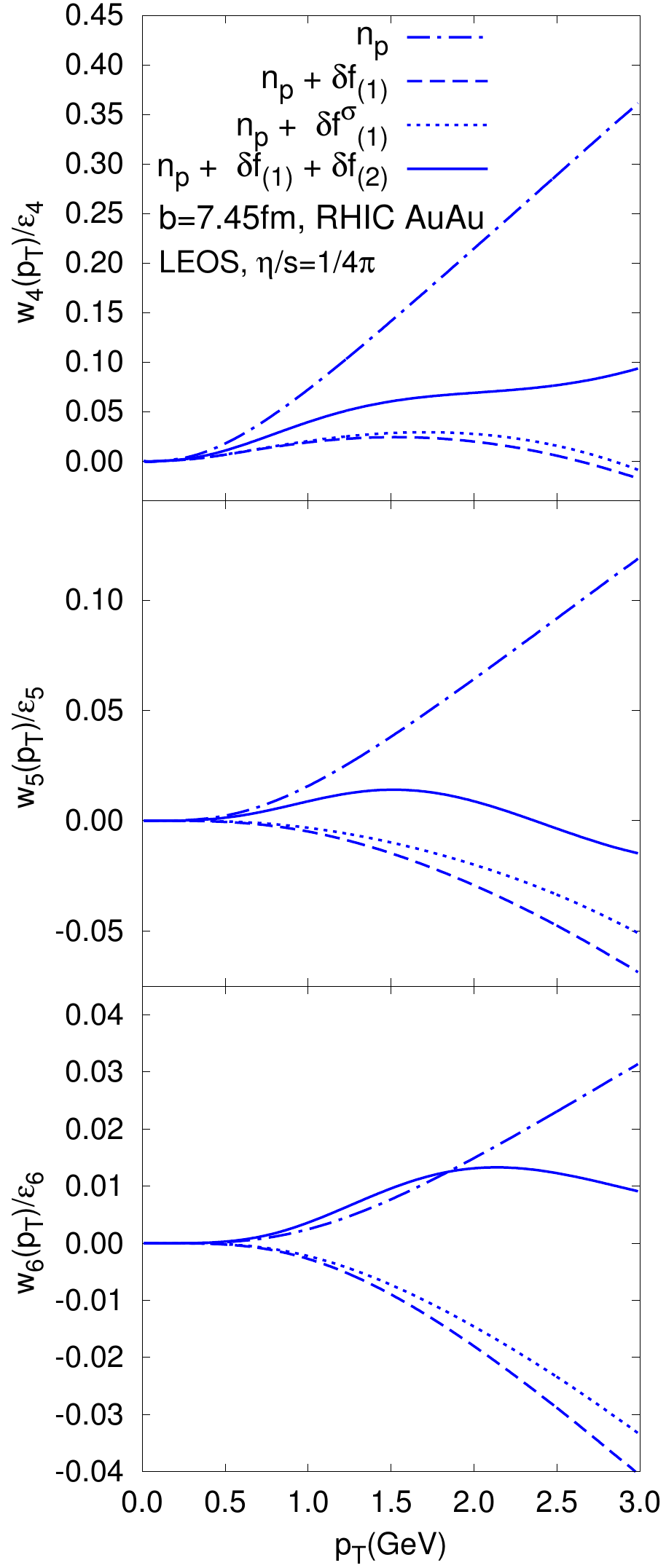} 
   \includegraphics[width=0.49\textwidth,height=7.6in]{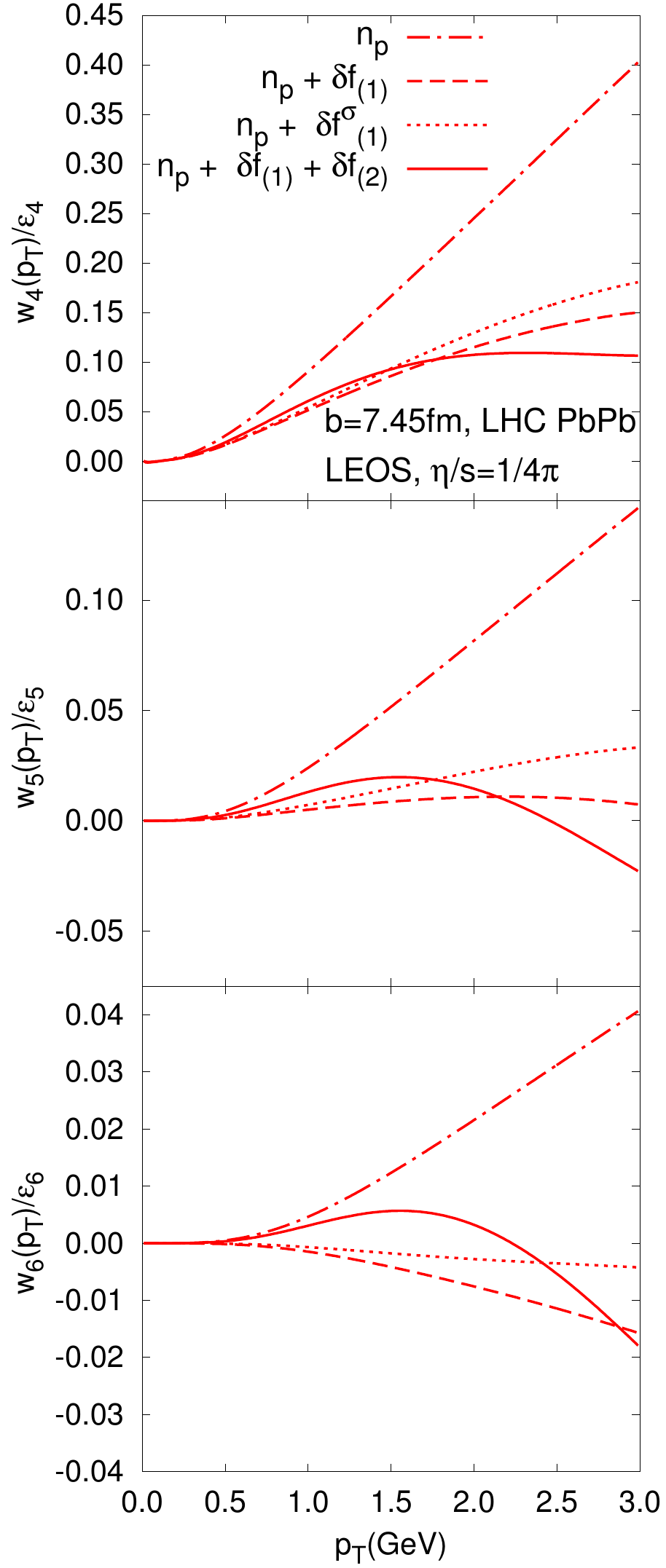} 
   \caption{Differential $w_n/\epsilon_n$  from various viscous
      hydrodynamic simulations  at $b=7.45\,{\rm fm}$ for RHIC and LHC initial conditions, and a lattice equation of state (LEOS) with $T_{\rm fo}=150\,{\rm MeV}$.
      Here 
      the $n_p$ curve shows the flow from second order hydrodynamics without
      the viscous correction to the distribution function;  $\delta f_{(1)}$
      and $\delta f_{(1)}+ \delta f_{(2)}$ show the flow with the viscous correction at first and second order respectively; and finally,
      the $\delta f_{(1)}^\sigma$ curve  uses $-\eta\sigma^{\mu\nu}$ instead of $\pi^{\mu\nu}$ in the first order viscous correction (see \Eq{df1sigma}).
\label{v4tov6}
}
\end{figure}

To summarize, in this section we have specified precisely the initial
conditions, the equation of state, the transport coeffiencts at  first and
second order, and the first and second order corrections to the distribution
functions. We have used this setup to compute the linear response coefficients
$w_n/\epsilon_n$ for RHIC and LHC initial conditions for the first six harmonics.
Our results are  displayed in \Fig{v1tov3} and \Fig{v4tov6}.  We will discuss the physics of these curves  in the next section.

\section{Discussion}
\label{discussion}

This work computed the second order viscous correction to the thermal
distribution function, $\dft$, and used this result to estimate the effect of
second order corrections on the harmonic spectrum. 
Our principle results are
shown in \Fig{v2ceos} for a conformal equation of state, and  \Fig{v1tov3}, and
\Fig{v4tov6}  for a lattice based equation of state.  First,
examine the $v_2$ curves for the conformal EOS shown in \Fig{v2ceos}.  The most
important remark is that  even for  a conformal equation of state, where the
expansion is most violent, the derivative expansion converges acceptably for
$p_T \lsim 1.5\, {\rm GeV}$, {\it i.e.} the second order correction is small
compared to the first order correction.  Not surprisingly, when a linear ansatz
is used for $\delta f$ (rather than the more popular quadratic ansatz) the
convergence  of the derivative expansion is improved  at high $p_T$.  Typically
in hydrodynamic simulations of heavy ion collisions, the strictly first order
$\delta f_{(1)}^{\sigma}$  is replaced by the $\delta f_{(1)}$ which
incorporates some, but not all, second order terms\footnote{As discussed above,
$\delta f_{(1)}$ uses $\pi^{\mu\nu}$ in place of  $-\eta \sigma^{\mu\nu}$  when
calculating the first order  correction.}.   Examining \Fig{v2ceos}, and also \Fig{v1tov3}
and \Fig{v4tov6},  we see that, while the sign of the  second order  correction
is correctly reproduced by this incomplete treatment, the magnitude of the  correction is
generally significantly underestimated,  and the $p_T$ dependence of the second
order correction is qualitatively wrong.   

Most of these observations  remain true for  the more realistic lattice
equation of state shown in \Fig{v1tov3} and \Fig{v4tov6}.  Generally, 
second order corrections are quite small for the first three harmonics, $v_1$ to $v_3$, and become increasingly important as the harmonic number increases. 
Indeed, for $v_6$ at RHIC and the LHC, 
the second order viscous correction is of order one, and the hydrodynamic
estimate can no longer  be trusted.  It is also instructive  to note that the sign of the second order viscous correction for $p_T \lsim 1.5\,{\rm GeV}$ is 
positive, {\it i.e.}  second order corrections bring the $v_n(p_T)$ curves closer to the ideal results.  Generally, when the first order correction, becomes
so large as to make $w_n$  negative\footnote{$v_n$ is positive by definition, see \Eq{definition}. Negative values of $w_n$ indicate that the flow angle $\Psi_n(p_T)$ is anti-aligned with the particpant angle plane, $\Phi_n$.},
the second order correction  conspires to  keep $w_n$  positive.
When  constraining 
$\eta/s$ with hydrodynamic simulations, the  second order corrections are most important for $v_4$ and $v_5$. 
Indeed,  at RHIC these corrections are quite important for $v_5$ even
in central collisions.  

For a practical perspective, using $\dft$ in a hydrodynamic simulation is  not particularly more difficult than using $\dfo$, and Eq.~\ref{f2start} can be readily implemented in most hydro codes. The functional form of $\dft$ and its magnitude is 
about as well constrained as $\dfo$,  and consistency with the second
order hydrodynamic evolution would seem to mandate its use.  At very least 
$\dft$  should be taken into consideration when estimating the 
uncertainty in the $\eta/s$ extracted from heavy ion collisions. Finally,
when trying to use hydrodynamics in very small systems such as proton-nucleus
collisions at RHIC and the LHC 
\cite{CMS:2012qk,Abelev:2012ola,Aad:2012gla,Bozek:2012gr,Adare:2013piz,Shuryak:2013ke,Bzdak:2013zma},  second order corrections to $\delta f$ should be used in order to monitor the convergence of the gradient expansion.

\vspace{\baselineskip}
\noindent{ \bf Acknowledgments:} \\
{}\\
We thank Ulrich Heinz for emphasizing that $v_n$ is positive by definition.
D.~Teaney is a RIKEN-RBRC fellow.  This work was supported in part by the
Sloan Foundation and by the Department of Energy, DE-FG-02-08ER4154.

\appendix

\section{Tensor decomposition of  $\dft$} 
\label{algebra}

The goal of this appendix is to compute $\dft$
and to record how this results transforms under rotations
in the local rest frame. Our starting point is Eqs.~(\ref{f2start}) 
which we rewrite in terms of irreducible tensors under rotations in the local rest frame.

A systematic strategy decomposes
all derivatives into temporal and spatial components
\st
 \partial_{\mu} = -U_{\mu} D + \nabla_{\mu} \, , 
\stp
where the spatial component $\nabla_{\mu}$ is orthogonal to $U_{\mu}$.
When  differentiating a quantity that is already first order ({\it i.e.}
$P^{\mu} \partial_{\mu} \delta f_{(1)}^{\sigma}$), we can use the lowest order conformal
equations of motion to rewrite time derivatives in terms of spatial derivatives 
\begin{align}
   D\ln T =&  - c_s^2 \nabla \cdot U \, , \\
   DU_{\mu} =&  -\nabla_{\mu} \ln T\, ,
\end{align}
where $c_s^2 = (e + \pr)/(Tc_v) =1/(d-1)$.
Finally, the resulting tensors  can be decomposed into symmetric, traceless,
and spatial tensors as in \Eq{tensor1}, which
transform irreducibly under rotations in the local rest frame.
To illustrate the procedure, 
we record the decomposition of $D\sigma_{\alpha\beta}$
\begin{align}
   D\sigma_{\alpha\beta} = D(\sigma^{\mu\nu}\Delta_{\mu\alpha} \Delta_{\nu\beta} ) =&
\Delta_{\mu\alpha} \Delta_{\nu\beta} D\sigma^{\mu\nu} + \sigma^{\mu\nu} D(\Delta_{\mu\alpha} \Delta_{\nu\beta} ) \, , \\
 =& \llangle D\sigma_{\alpha\beta} \rrangle   
- (\sigma^{\mu}_{\;\beta} u_{\alpha} \nabla_\mu \ln T +  u_{\beta} \sigma_{\;\alpha}^\nu \nabla_{\nu} \ln T) \, .
\end{align}
Similarly, the symmetrized spatial tensor $\left\{\nabla_{\mu}\sigma_{\alpha\beta} \right\}_{\rm sym}$
that arises when differentiating $\delta f_{(1)}^{\sigma}$
is decomposed as 
\begin{multline}
    \left\{\nabla_{\mu}\sigma_{\alpha\beta} \right\}_{\rm sym}
    = 
\llangle \nabla_{\mu} \sigma_{\alpha\beta} \rrangle +  \Big\{\frac{2}{d+1} 
 \Delta_{\mu\alpha} \nabla_\gamma \sigma^{\gamma}_{\;\beta}    
  + u_{\alpha} \llangle \sigma^{\rho}_{\;\beta}\sigma_{\mu\rho} \rrangle   \\
  + u_{\alpha} \frac{\Delta_{\mu\beta}}{d-1} \sigma^2   
    + 2 u_{\alpha} \llangle \sigma^{\rho}_{\;\beta}\Omega_{\mu\rho} \rrangle + 
 2 u_{\alpha} \frac{\sigma_{\mu\beta}}{d-1} \nabla \cdot U  
    \Big \}_{\rm sym} \, ,
 \end{multline}
where we  have used
\st
 \nabla_{\mu} u_{\rho} = \frac{1}{2} \sigma_{\mu\rho} + \Omega_{\mu\rho} + \frac{\Delta_{\mu\rho}}{d-1} \nabla \cdot U  \, .
\stp
Finally, we note that 
\st
\label{hydroid}
\llangle\partial_\lambda \pi^{\lambda}_{\;\mu}\rrangle  = \Delta_{\mu\lambda_2}
\,  \partial_{\lambda_{1} } \pi^{\lambda_1 \lambda_2}   
    =   -\eta \left[ (d-2) \llangle \sigma_{\mu \lambda} \nabla^{\lambda} \ln T \rrangle  + \llangle \nabla_{\lambda} \sigma^{\lambda}_{\; \mu} \rrangle \right]   \, , 
\stp
where we have used the first order expression,
$\pi^{\mu\nu} = - \eta \sigma^{\mu\nu}$,  
the conformal temperature dependence of $\eta \propto T^{d-1}$, and
the lowest order equations of motion. 

With this automated set of steps, we start with
\Eq{f2start} and place 
$\delta f_{(2)}^{\sigma}$ into its canonical form
\begin{align}
   \delta f_{(2)}^{\sigma}= & 
   \chi_{1p} \, \frac{P^{\mu_1}P^{\mu_2}P^{\mu_3}P^{\mu_4}}{T^6}  \llangle \sigma_{\mu_1\mu_2} \sigma_{\mu_3 \mu_4} \rrangle      
   + \chi_{2p} \, \frac{P^{\mu_1} P^{\mu_2} P^{\mu_3}}{T^5} \left[  \llangle \nabla_{\mu_1} \sigma_{\mu_2 \mu_3}  \rrangle  - 3 \llangle \sigma_{\mu_1\mu_2} \nabla_{\mu_3} \ln T \rrangle \right]   \nonumber \\
   & \qquad \qquad + 
   \left( \chi_{1p} \frac{4\bar p^2}{d+3} -  \chi_{2p}  \bar E_p \right)  \frac{P^{\mu_2} P^{\mu_1}  }{T^4}
 \llangle \sigma^\lambda_{\;\mu_2}\sigma_{\mu_1\lambda} \rrangle  \nonumber  \\
 & \qquad \qquad +   
\chi_{2p} \bar E_p\,  \frac{P^{\mu_2} P^{\mu_1} }{T^4} \Big[
     \llangle D\sigma_{\mu_2\mu_1}  \rrangle + \frac{\sigma_{\mu_2\mu_1}}{d-1} \nabla \cdot U 
- 2\llangle
\sigma^{\lambda}_{\;\mu_2}\Omega_{\mu_1\lambda} \rrangle   
\Big] \nonumber  \\  
 & \qquad \qquad + \xi_{3p} \frac{P^{\mu_2}}{T^3} \left[ 
-\llangle \nabla_{\lambda} \sigma^{\lambda}_{\;\mu_2} \rrangle   - 
(d-2) \llangle \sigma_{\mu_2\lambda} \nabla^{\lambda} \ln T \rrangle \right]  
 + \frac{\xi_{4p}}{T^2} \sigma^2    \, ,
\end{align}
where the functions $\chi_{0p}$, $\chi_{1p}$, $\chi_{2p}$ and $\xi_{3p}$
and $\xi_{4p}$ are recorded in the text, \Eq{xieq}.

In this form it is easy to integrate over the the phase space to determine
the viscous stress
\st
\pi^{\mu\nu} = \pi^{\mu\nu}_{(1)} +  \pi^{\mu\nu}_{(2) }  = \int_p \frac{P^{\mu} P^{\nu} }{P^0} \left( \delta f_{(1)}^{\sigma} + \delta f_{(2)}^{\sigma}  \right) \, ,
\stp
where $\pi^{\mu\nu}_{(1)}$ and $\pi^{\mu\nu}_{(2)}$ are given 
by static form of the constituent relation \Eq{constituent}.
Rotational invariance in the rest frame 
reduces these tensor integrals, {\it e.g.}
\st
\int_\p  \chi_{0p} \frac{P^{\mu_1} P^{\mu_2} P^{\mu_3} P^{\mu_4}}{P^0}  \llangle O_{\mu_3 \mu_3} \rrangle  = \left[ \frac{2}{(d-1)(d+1) } \int_p \chi_{0p} \frac{p^4}{E_\p} \right]
    \llangle O^{\mu_1\mu_2}  \rrangle, 
\stp
yielding the equations for the transport coefficients written 
in the text,  Eqs.~(\ref{sheareq}),~(\ref{l1eq}), and (\ref{nteq}). 
In addition, we see that independent of the collision integral one 
finds the kinetic theory expectations identified in \Ref{York:2008rr}
\st
\lambda_2 = -2 \eta \tau_\pi\, ,  \qquad \mbox{and} \qquad \lambda_3 = 0  \, .
\stp

Finally, in presenting these results in the text, and in implementing 
the results in a realistic hydrodynamic simulation, we have 
used the dynamic form of second order hydrodynamics, where $\pi^{\mu\nu}$
is treated as a dynamic variable.
This choice amounts to using $-\pi_{\mu\nu}/\eta$
in place of $\sigma_{\mu\nu}$.  In $\delta f_{(2)}^{\sigma}$  this reparametrizations yields the replacements:
\begin{align}
     \llangle D\sigma_{\mu_1\mu_2}  \rrangle + \frac{\sigma_{\mu_1\mu_2}}{d-1} \nabla \cdot U 
- 2\llangle
\sigma^{\lambda}_{\;\mu_1}\Omega_{\mu_2\lambda} \rrangle   &\rightarrow \frac{1}{\eta\tau_\pi}[\pi_{\mu_1\mu_2}  + \eta\sigma_{\mu_1\mu_2} ]  - \frac{\lambda_1}{\eta\tau_\pi} \frac{1}{\eta^2} \llangle \pi_{\;\;\mu_1}^{\lambda} \pi_{\mu_2\lambda} \rrangle \,   ,  \\
\llangle \nabla_{\mu_1} \sigma_{\mu_2 \mu_3}  \rrangle  - 3 \llangle \sigma_{\mu_1\mu_2} \nabla_{\mu_3} \ln T \rrangle  
&\rightarrow  \frac{1}{\eta} \left[  (d+2) \llangle \pi_{\mu_1\mu_2} \nabla_{\mu_3} \ln T \rrangle 
   - \llangle \nabla_{\mu_1} \pi_{\mu_2 \mu_3}  \rrangle   \right] \, ,
\end{align}
and \Eq{hydroid}.  
In addition, when replacing $\sigma_{\mu\nu}$ with $-\pi_{\mu\nu}/\eta$ in the first order result, 
the difference $\frac{1}{\eta} (\pi_{\mu\nu} + \eta \sigma_{\mu\nu})$ 
must be appendend to the second order result -- see \Eq{sigtopi}.
The full result for $\delta f_{(1)}$ and $\delta f_{(2)}$ is given
in Eqs.~(\ref{f1}) and (\ref{f2}) respectively. 

\bibliography{df2bib}

\end{document}